\documentclass[10pt,twocolumn,landscape,a4paper]{article}

\font\mybb=msbm10 at 10pt
\def\bb#1{\hbox{\mybb#1}}

\def\R {\bb{R}}
\begin{document}
\begin{titlepage}

\begin{flushright}
CERN-TH/95--347\\
McGill/96--01\\
{\bf hep-th/9512178}\\
December $21$st, 1995\\ 
(Revised January $31$st, $1996$)
\end{flushright}

\begin{center}

\baselineskip25pt
{\LARGE {\bf A NON--SUPERSYMMETRIC DYONIC EXTREME REISSNER--NORDSTR\"{O}M
BLACK HOLE}}

\vspace{1cm}

{\large{\bf Ramzi R.~Khuri}${}^{a,b}$
\footnote{E-mail address: {\tt khuri@nxth04.cern.ch}}
        {\bf and Tom\'as Ort\'{\i}n}${}^{a}$
\footnote{E-mail address: {\tt tomas@surya20.cern.ch}}\\

\vspace{.5cm}

${}^{a}${\it C.E.R.N.~Theory Division}\\
{\it CH--1211, Gen\`eve 23, Switzerland}\\

\vspace{.5cm}

${}^{b}${\it McGill University}\\
{\it Physics Department}\\
{\it Montreal, PQ, H3A 2T8 Canada}\\
}
\end{center}

\vspace{1.5cm}

\begin{abstract}

We present a dyonic embedding of the extreme Reissner-Nordstr\"om black
hole in $N=4$ and $N=8$ supergravity that breaks all supersymmetries.

\end{abstract}

\vspace{1cm}

\begin{flushleft}
CERN-TH/95--347\\
McGill/96--01
\end{flushleft}

\end{titlepage}

\newpage

\baselineskip11pt

\pagestyle{plain}


It has recently been shown \cite{kn:KO} that all possible embeddings of
purely electric or purely magnetic four-dimensional extreme dilaton
black holes in $N=8$ supergravity must be related by duality symmetries
of the $N=8$ supergravity theory (``$U$~duality'' \cite{kn:HT}) and
therefore must have the same number of unbroken supersymmetries in
$N=8$.  As explained in Ref.~\cite{kn:KO}, the same statement is not
true in the context of the $N=4$ theory and other truncations of the
$N=8$ theory with fewer supersymmetries because some of the $U$~duality
transformations that relate the embeddings (which are also strictly
$N=4$ embeddings) are {\it not} duality symmetries of the $N=4$ theory,
and change the number of unbroken $N=4$ supersymmetries.

A well-known example is that of the $a=1$ extreme dilaton black hole
solution of the action \cite{kn:GGM}

\begin{equation}
S^{(a)}= {\textstyle\frac{1}{2}}\int d^{4}x\ \sqrt{|\tilde{g}|}
\left[-\tilde{R} -2(\partial\varphi)^{2}
+{\textstyle\frac{1}{2}}e^{-2a\varphi}F^{2}\right]\, .
\label{eq:action}
\end{equation}

\noindent When Garfinkle, Horowitz and Strominger rediscovered this
solution in the context of string theory \cite{kn:GHS}, they made the
claim that, in spite of the fact that the solution saturates a
Bogomol'nyi-type bound, it is not supersymmetric because the photino
supersymmetry transformation law was always different from zero for any
supersymmetry parameter.  In other words, when the vector field
appearing in the action considered by the authors was taken to be a
matter vector field belonging to a vector supermultiplet, the
supersymmetry variation of the spinor present in this supermultiplet
could not be made to vanish, and the solution was not supersymmetric.

Shortly afterwards \cite{kn:KLOPP} it was discovered that, when the
vector field is considered as one of the six vector fields present in
the $N=4,D=4$ supergravity multiplet, there exists a choice of local
supersymmetry parameter such that the supersymmetry transformation laws
of the four gravitini and dilatini of this supermultiplet vanish; {\it
i.e.} there exists a Killing spinor such that the supersymmetry
variation vanishes and the extreme dilaton black hole is supersymmetric.

The choice of considering the vector field as a matter or as a
supergravity vector field corresponds to a different choice of {\it
embedding} of the same four-dimensional solution in the ten-dimensional
$N=1$ supergravity theory which corresponds to the $N=4$ supergravity
theory plus six vector supermultiplets in four dimensions upon
dimensional reduction \cite{kn:C}.

This seemingly paradoxical situation in which the same solution is or
is not supersymmetric depending on the embedding in $N=1,D=10$
supergravity chosen is resolved by further embedding into any of the
$N=2,D=10$ theories \cite{kn:KO}. The six matter vector fields of the
$N=4,D=4$ theory become supergravity vector fields of the $N=8,D=4$
theory that one obtains by dimensional reduction of $N=2,D=10$
supergravity. One finds that the two embeddings proposed in
Refs.~\cite{kn:GHS,kn:KLOPP} have the same number of $N=8$
supersymmetries unbroken (namely two).

The reason for this result (and analogous results for the other values
of $a$ that occur in $N=8,D=4$ supergravity) is that all the
different embeddings of the same solution are related by $U$~duality
symmetries, which preserve all the $N=8,D=4$ supersymmetries.  These
symmetries are not always duality symmetries of the $N=4,D=4$ theory,
and, therefore, even in the cases in which the $U$~duality symmetry
takes us from an $N=4,D=4$ solution to another $N=4,D=4$ solution, the
number of unbroken $N=4,D=4$ supersymmetries is generally not preserved.

In this letter we present a new embedding of the $a=0$
extreme black hole ({\it i.e.}~the extreme Reissner-Nordstr\"om (ERN)
black hole) that has no unbroken supersymmetries in $N=8,D=4$
supergravity.  Since all the other known embeddings have one $N=8,D=4$
unbroken supersymmetry, this embedding cannot be related to the known
embeddings by any $U$~duality symmetry.  It is also unique
because it is the only embedding of dyonic nature of any extreme dilaton
black hole known (with just one vector field and no axion field).

Since the embedding is determined by the identification of the $N=8,D=4$
fields with the fields occurring in the action (\ref{eq:action}) we
first need to present the $N=8,D=4$ action. In fact, this embedding is
also an embedding in the $N=4,D=4$ (plus six vector supermultiplets)
truncation of the $N=8,D=4$ theory, whose action is \cite{kn:MS}

\begin{eqnarray}
S & = & {\textstyle\frac{1}{2}} \int d^{4}x\ \sqrt{|g|}\ e^{-2\phi}
\left\{ -R +4(\partial\phi)^{2} -{\textstyle\frac{3}{4}}H^{2} \right.
\nonumber \\
& &
\nonumber \\
& &
+{\textstyle\frac{1}{4}}\left[\partial G_{mn} \partial G^{mn} -
G^{mn} G^{pq} \partial B_{mp} \partial B_{nq}\right]
\nonumber \\
& &
\nonumber \\
& &
\left.
-{\textstyle\frac{1}{4}} \left[ G_{mn}F^{(1)m}F^{(1)n}
+G^{mn}{\cal F}_{m}{\cal F}_{n} \right]
\right\}\, .
\label{eq:n4action}
\end{eqnarray}

It can be easily checked that the equations of motion of this theory
reduce to those of the theory (\ref{eq:action}) with $a=0$ with the
identifications ({\it i.e.}~the embedding)

\begin{equation}
\tilde{g}_{\mu\nu}=g_{\mu\nu}\, ,
\hspace{.5cm}
F^{(1)1}{}_{\mu\nu}=F_{\mu\nu}\pm {}^{\star}F_{\mu\nu}\, ,
\hspace{.5cm}
\varphi= {\rm constant}\, .
\end{equation}

\noindent where $F$ is either purely electric or purely magnetic, so
$F^{(1)1}{}_{\mu\nu}{}^{\star}F^{(1)1}{}_{\mu\nu}=0$.  Then, any solution with
constant
scalar and purely electric or magnetic vector field of the
Einstein-Maxwell-scalar theory can be considered as a solution of the
above $N=4,D=4$ theory, and, henceforth, of the $N=8,D=4$ theory.  In
particular, one can take the purely electric ERN solution

\begin{equation}
\left\{
\begin{array}{rcl}
d\tilde{s}^{2} & = & V^{-2} dt^{2}
-V^{+2}d\vec{x}^{2}\, ,
\\
& &
\\
F_{t\underline{i}} & = & - \sqrt{2}\ n\
\partial_{\underline{i}}V^{-1}\, ,
\end{array}
\right.
\label{eq:ern}
\end{equation}

\noindent where $V(\vec{x})$ is a harmonic function in three-dimensional
Euclidean space $\partial_{\underline{i}} \partial_{\underline{i}}V =0$

\begin{equation}
V(\vec{x})= 1 +\sum_{i}\frac{M_{i}}{|\vec{x}-\vec{x}_{i}|}\, ,
\hspace{.5cm}M_{i}\geq 0\hspace{.2cm}\forall i=1,2,3\, ,
\end{equation}

\noindent and $n=\pm 1$ gives the sign of all the electric charges.

The advantage of the form (\ref{eq:n4action}) of the $N=4,D=4$ theory is
that the relation between the four-dimensional and the ten-dimensional
fields is simple and known \cite{kn:MS,kn:KO}.  This allows us to
rewrite the dyonic ERN solution (\ref{eq:ern}) as a solution of $N=1,D=10$
supergravity (and, therefore, of the two $N=2,D=10$ theories):

\begin{equation}
\left\{
\begin{array}{rcl}
d\hat{s}^{2} & = & V^{-2}dt^{2} -V^{2}d\vec{x}^{2}
-\left[ dx^{\underline{4}}
+\sqrt{2}n\ \left( V^{-1}dt \pm V_{\underline{i}}
dx^{\underline{i}}\right) \right]^{2}\\
& & \\
& &
-dx^{\underline{I}} dx^{\underline{I}}\, ,
\hspace{1cm} I=5,\ldots,9\, .\\
& & \\
\hat{B} & = & \hat{\phi} = 0\, ,\\
\end{array}
\right.
\end{equation}

\noindent where the $V_{\underline{i}}$'s are functions satisfying

\begin{equation}
\partial_{[\underline{i}}V_{\underline{j}]} =
{\textstyle\frac{1}{2}}\epsilon_{ijk}\partial_{\underline{k}}V\, ,
\end{equation}

\noindent and whose (local) existence is ensured by the harmonicity of
$V$.

This configuration is purely gravitational in ten dimensions.  (One can
directly check that it is a solution of ten-dimensional $N=1$
supergravity {\it i.e.}~the metric above is Ricci-flat\footnote{We thank
H.H.~Soleng for help with the {\sl Mathematica} package {\sc CARTAN}.}.)
This implies that both torsionful spin connections $\hat{\Omega}^{(\pm)}
=\hat{\omega} \mp{\textstyle\frac{3}{2}} \hat{H}$ are equal to the
(torsionless) spin connection $\hat{\omega}$ whose (tangent space)
components are

\begin{equation}
\begin{array}{rcl}
\hat{\omega}_{0}{}^{0i} & = &
-V^{-2}\partial_{\underline{i}}V\, ,\\
& & \\
\hat{\omega}_{4}{}^{0i} =\hat{\omega}_{i}{}^{04}=
\hat{\omega}_{0}{}^{i4} & = &
{\textstyle\frac{1}{\sqrt{2}}}n\ V^{-2}\partial_{\underline{i}}V\, , \\
& & \\
\hat{\omega}_{4}{}^{ij}= \hat{\omega}_{j}{}^{i4} & = & \mp \sqrt{2} n\
V^{-2} \partial_{[\underline{i}} V_{\underline{j}]}\, , \\
& & \\
\hat{\omega}_{k}{}^{ij} & = &
2V^{-2}\delta_{k[i}\partial_{\underline{j}]} V\, .
\end{array}
\label{kn:spincon}
\end{equation}

The vanishing of the axion and dilaton fields allows us to check the
whole $N=2,D=10$ supersymmetry rules by checking just the
gravitini rules \cite{kn:KO}:

\begin{equation}
\delta_{\epsilon} \hat{\psi}^{(\pm)}_{\hat{a}}=
\hat{\nabla}_{\hat{a}} \hat{\epsilon}^{(\pm)}\, ,
\end{equation}

\noindent where the superscript ${}^{(\pm)}$ of the spinors indicates
their ten-dimensional chirality and $\hat{\nabla}_{\hat{a}}$ is the
covariant derivative associated to the usual spin connection whose
components can be read in Eqs.~(\ref{kn:spincon}).

Now, if this configuration is going to have any four-dimensional
unbroken supersymmetry, the corresponding Killing spinor should depend
neither on the six compact coordinates
$x^{\underline{4}},\ldots,x^{\underline{9}}$ nor on the time coordinate:

\begin{equation}
\partial_{t}\hat{\epsilon}^{(\pm)}=
\partial_{\underline{4}}\hat{\epsilon}^{(\pm)}= \ldots =
\partial_{\underline{9}}\hat{\epsilon}^{(\pm)}= 0\, ,
\end{equation}

\noindent and, substituting these conditions and the spin connection
components into the zero component of the Killing spinor equation, we get
the condition

\begin{equation}
\delta\hat{\psi}_{0}^{(\pm)}=0\,\,\, \Rightarrow\,\,\,
\hat{\gamma}_{0}\hat{\gamma}_{4}
\hat{\epsilon}^{(\pm)}= {\textstyle\frac{1}{\sqrt{2}}}n\
\hat{\epsilon}^{(\pm)} \, .
\end{equation}

\noindent The matrix $\hat{\gamma}_{0}\hat{\gamma}_{4}$ squares to one,
which means that its only eigenvalues are $\pm 1$, and from this it
follows that $\hat{\epsilon}^{(\pm)}=0$ and all the supersymmetries are
broken.  Had the coefficient on the r.h.s.~been $1$ instead of
$1/\sqrt{2}$, the above condition could be fulfilled for some spinors.
The origin of this $1/\sqrt{2}$ factor can be traced to a continuous
electric-magnetic duality rotation of a purely electric or magnetic ERN
black hole that transforms it into a dyon with electric and magnetic
charges of equal absolute value.

This result is in direct contrast to our findings for purely electric or
purely magnetic ERN ($a=0$) black holes \cite{kn:KO}, namely, that one
supersymmetry is always preserved in $N=8$ (but not in $N=4$).  One
obvious difference is that the solution considered here is dyonic.
Observe, though, that must be at least one supersymmetric embedding of
the dyonic ERN black hole which corresponds to the truncation from $N=8$
to $N=2$ supergravity.  In the $N=2,D=4$ theory, the purely electric or
magnetic ERN black holes are supersymmetric and the electric-magnetic
duality of this theory respects supersymmetry (see, for instance, the
second lecture in Ref.~\cite{kn:GWG}).  The supersymmetric embedding
in the $N=8$ theory of the ERN dyonic black hole must have
more than one non-vanishing vector field.

An intriguing possible explanation of the nonsupersymmetric nature of
the dyonic black hole is the following: just as solutions which
preserved different numbers of supersymmetries in $N=4$ but the same
number in $N=8$ were argued to be unrelated by $U$~duality
transformations of $N=4$ but connected by $U$~duality transformations in
$N=8$, the question can be raised as to whether the present embedding is
related to another supersymmetric embedding (for instance those of its
purely electric or magnetic counterparts) by a duality transformation
which is not a $U$~duality transformation.  That would lead us to
speculate about the existence of a supergravity theory {\it larger}
than $N=8$ of which that duality would be a symmetry, just as
$C$~duality is a symmetry in the $N=8$ theory but not in its $N=4$
truncations.

Higher supergravity theories (higher than $N=1,D=11$) are usually
considered ``unphysical'' because they must necessarily contain
particles with helicities higher than two \cite{kn:N} and then
consistency problems arise in the quantum field theory of these
particles. As pointed out in Ref.~\cite{kn:S}, this could just be a
prejudice reflecting the current status of local field theory.
However, there are other serious problems in higher supergravity
theories, such as the existence of more than one metric.

Still, even if this higher supergravity theory were physically
unacceptable, one could still conceive that such an unphysical theory
exists and is such that {\it it has a consistent physically acceptable
truncation}. We have in mind the analogues of the truncations of the
type~II theories that render $N=1,D=10$ supergravity: one consistently
truncates a set of fields (the R-R fields) and gets another supergravity
theory. In the higher supergravity case, all the unwanted fields (higher
helicities and further metrics, for instance) should disappear in this
truncation. The result should be a known, physically acceptable,
supergravity theory.

It is tempting to take this speculation a bit further and try to
identify the truncated theory.  A good candidate could be the type~IIB
theory, and a possible scenario could be the following: the higher
supergravity theory could be non-chiral $N=4,D=10$, and there would be a
chiral truncation in which all unphysical fields disappear and one
chiral sector also disappears.  Since the chirality of the type~IIB
theory is related to the self-or anti-self-duality of the five-form, and
both chiralities would be present in this higher supergravity, the whole
of the five form would also be present.  Although it is highly
speculative, this scenario is particularly attractive because the higher
supergravity theory could be obtained from higher dimensions and the
origin of the type~IIB could be ``explained'' in a fashion analogous to
the type~IIA.  On the other hand, the higher supergravity theory could
encompass all known supergravities (it could have a truncation giving
$N=1,D=11$), in the spirit of the $M$~theory \cite{kn:M}.  Observe that
type~IIB supergravity has as global symmetry group $GL(2,\R)$, and this
would be consistent with its higher-dimensional origin \cite{kn:BJO}.

Perhaps there are other explanations, possibly in terms of 
the bound state picture of extremal black holes \cite{kn:R},
that make the above speculation unnecessary. Furthermore,
non-supersymmetric ERN black holes corresponding to 
massless states and arising from
a more general ansatz \cite{kn:CY4P} were found in Refs.~\cite{kn:CYSS},
while large classes of non-supersymmetric ERN black holes for generic choice
of Page charge were found in Refs.~\cite{kn:LPSS}.  
The abundance of these solutions may even render an explanation of our 
results unnecessary
to begin with, if we accept that identical four-dimensional metrics can
correspond to bound states which are different and inequivalent under duality,
and have different supersymmetry properties, a proposition which we still
find rather mysterious.


\section*{Acknowledgements}

T.O.~would like to thank M.M.~Fern\'andez for her support and
encouragement.  R.K.~would like to thank the East Carolina Department of
Mathematics for their hospitality.  R.K.~was supported by a World
Laboratory Fellowship.


\end{document}